\documentclass[prb,twocolumn,superscriptaddress]{revtex4}
\usepackage{amsmath,amssymb}
\usepackage{graphicx}

\begin{document}

\title{Ballistic (precessional) contribution to the conventional magnetic switching}

\author{Ya.\ B. Bazaliy}
 \email{yar@physics.sc.edu}
 \affiliation{Department of Physics and Astronomy, University of South Carolina, Columbia, SC 29208, USA}
 \affiliation{Institute of Magnetism, National Academy of Science, Kyiv 03142, Ukraine}

\author{Andrzej Stankiewicz}
 \affiliation{NVE Corporation, Eden Prairie, MN 55344, USA}

\date{\today}

\begin{abstract}
We consider a magnetic moment with an easy axis anisotropy energy,
switched by an external field applied along this axis. Additional
small, time-independent bias field is applied perpendicular to the
axis. It is found that the magnet's switching time is a
non-monotonic function of the rate at which the field is swept from
``up'' to ``down''. Switching time exhibits a minimum at a
particular optimal sweep time. This unusual behavior is explained by
the admixture of a ballistic (precessional) rotation of the moment
caused by the perpendicular bias field in the presence of a variable
switching field. We derive analytic expressions for the optimal
switching time, and for the entire dependence of the switching time
on the field sweep time. The existence of the optimal field sweep
time has important implications for the optimization of magnetic
memory devices.
\end{abstract}

\maketitle

In conventional magnetic switching by an externally applied magnetic
field the moment performs many revolutions before switching to the
opposite direction. While being much slower than the ballistic
(precessional) switching \cite{back:1999,xiao:2006,wang:2007} which
is tested experimentally but not yet realized in applications,
conventional switching is used for magnetic recording in hard disk
drives and other devices. The speed at which the moments of magnetic
bits can be switched between the two easy directions has obvious
implications for the technology performance, setting the limit for
the information writing rate. Here we study the dependence of the
conventional magnetic switching time $\tau_m$ on the reversal time
of the writing head field. If the field is swept from ``up'' to
``down'' in a time $\tau_h$, the switching time will be a function
$\tau_m(\tau_h)$. It would seem natural to assume that $\tau_m$
decreases with decreasing $\tau_h$ and the fastest switching is
realized by an instantaneous flip of the field with $\tau_h = 0$.
However, it was found numerically by one of the authors
\cite{stankiewicz} that in the presence of a small perpendicular
bias field the function $\tau_m(\tau_h)$ is not monotonic and has a
minimum at an optimal sweep time $\tau_{h}^*$. Decreasing $\tau_h$
below the optimal value would be counterproductive in terms of the
technology performance. In this paper we provide analytic
approximations for the function $\tau_m(\tau_h)$ and the optimal
field sweep time $\tau_{h}^*$. We find that the nonmonotonic
behavior of $\tau_m(\tau_h)$ is a result of the admixture of a
``ballistic'' (or ``precessional'') switching induced by the
perpendicular bias field. Ballistic contribution is normally
quenched by the anisotropy, but here it is restored by the
time-dependence of the field during the rise time of the applied
step. The effect considered here is different from the decrease of
the switching time \cite{suess:2002} predicted for switching below
the Stoner-Wholfarth limit.\cite{he:1994,porter:1998} The latter
consists of the $\tau_m$ dependence on the amplitude of the field
step with an instantaneous rise time, while in our case the finite
rise time is essential. Conventional switching considered here is
achieved by a field step and does not require precisely timed pulses
of finite duration needed for a truly ballistic switching.
\cite{back:1999,xiao:2006,wang:2007} Also, the required field
magnitude is much smaller than in the ballistic case.

\begin{figure}[b]
\center
\includegraphics[width=0.4\textwidth]{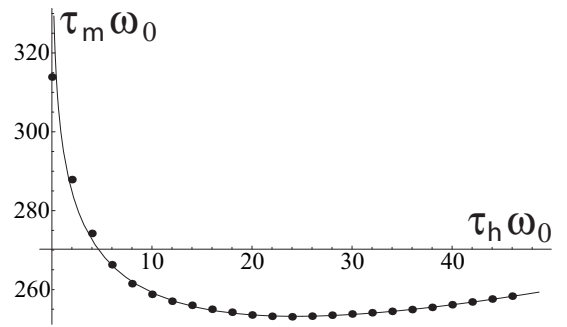}
\caption{Switching time $\tau_m$ as a function  of field sweep
time~$\tau_h$. Here $\alpha = 0.01$, $h_0 = 3.5 \ \omega_0$,
$h_{\perp} = 0.001 \ \omega_0$. The solid line is given by the
analytical expression (\ref{eq:tau_m_preview}).}
 \label{fig:tsw_numeric}
\end{figure}

Magnetic particles of nanometer size are single-domain
\cite{back:1999,thirion:2003} and can be described by the moment
${\bf M} = M_0 {\bf n}$, where ${\bf n}$ is a unit vector. We
consider a particle with an easy axis $\hat z$ and anisotropy energy
$E_a = -(1/2)K n_z^2$. The switching field ${\bf H} = H(t) \hat z$
is directed along the easy axis. For large enough field magnitudes,
$|H| > K/M_0$, only one equilibrium direction of $\bf M$ is stable.
A field applied exactly along the axis leads to a magnetization
switch only when some fluctuations of $\bf M$ are present. Following
Ref.~\onlinecite{stankiewicz} we introduce a small constant
perpendicular bias field ${\bf H}_{\perp} = H_{\perp} \hat x$,
$H_{\perp} \ll K/M_0$ to mimic the required fluctuations.

The dynamics of the moment are governed by the
Landau-Lifshitz-Gilbert (LLG) equation
$$
\dot{\bf M} = -\gamma \left[\frac{\partial E}{\partial {\bf M}}
\times {\bf M} \right] + \frac{\alpha}{M_0} [{\bf M} \times \dot{\bf
M}] \ ,
$$
where $\gamma$ is the gyromagnetic ratio, $E = E_a - ({\bf H} + {\bf
H}_{\perp})\cdot {\bf M}$ is the total magnetic energy, and $\alpha
\ll 1$ is the Gilbert damping constant. The field sweep is assumed
to be linear in time and given by the expressions $H(t) = + H_0$ for
$t < 0$, $H(t) = H_0 ( 1 - 2 t/\tau_h )$ for $0 < t < \tau_h$, and
$H(t) = - H_0$ for $t > \tau_h$. As the field is swept from positive
to negative values, the up-equilibrium disappears and the magnetic
moment starts to move towards the down-equilibrium along a spiral
trajectory. The final approach to the down-equilibrium in
exponential. To define a finite switching time we have to introduce
a provisional cut-off angle $\theta_{sw}$ and calculate the time it
takes to reach it. The extra time needed to cover the remaining
distance does not depend on $\tau_h$ because that part of  the
motion happens at a constant field $H = -H_0$. In accord with
Refs.~\onlinecite{stankiewicz, suess:2002, he:1994, porter:1998} we
choose $\theta_{sw} = \pi/2$.

\begin{figure}[t]
\center
\includegraphics[width=0.48\textwidth]{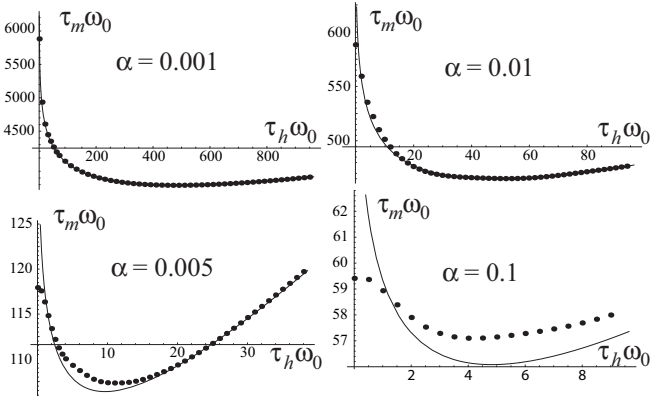}
\caption{Dependencies $\tau_m(\tau_h)$ calculated for $h_0 = 2.2 \
\omega_0$, $h_{\perp} = 0.001 \ \omega_0$, and variable $\alpha$
indicated on each panel. As $\alpha$ increases, the theoretical fit
gets poorer due to the violation of the strong inequality
(\ref{eq:alpha_requirement}).}
 \label{fig:alpha_family}
\end{figure}

The LLG equation can be easily solved numerically and the switching
time dependence $\tau_m(\tau_{h})$ can be obtained.
Fig.~\ref{fig:tsw_numeric} shows the results of such modeling for a
particular parameter set. The minimum of $\tau_m$ is clearly
observed.

We have derived an approximate analytic expression for the switching
time. Denoting $h_0 = \gamma H_0$, $h_{\perp} = \gamma H_{\perp}$,
and $\omega_0 = \gamma K/M_0$ we find
\begin{eqnarray}\label{eq:tau_m_preview}
\tau_m & \approx & \frac{3 h_0 + \omega_0}{4 h_0} \tau_h +
\frac{\ln[h_0/\pi h_{\perp}^2 \tau_h]}{2 \alpha(h_0 -\omega_0)} +
\tau_R \ ,
\end{eqnarray}
where $\tau_R$ is a part independent of $\tau_h$ and $h_{\perp}$
\begin{eqnarray*}
 \tau_R &=&  \frac{1}{2\alpha} \left\{
 \frac{1}{h_0 - \omega_0}
 \ln \left(
   \frac{2(h_0 - \omega_0)}{h_0}
 \right) \right. -
 \\ \label{eq:def_R}
 &&
 - \left.
 \frac{1}{h_0 + \omega_0}
 \ln \left(
   \frac{h_0 - \omega_0}{2 h_0}
 \right)
 \right\} \ .
\end{eqnarray*}
Formula (\ref{eq:tau_m_preview}) is the first main result of our
paper. As one can see in Fig.~\ref{fig:tsw_numeric}, it reproduces
the function $\tau_m(\tau_h)$ quite well.

Approximation (\ref{eq:tau_m_preview}) requires small $h_{\perp}$
and $\alpha$. For a given $h_{\perp}$  it is valid in the interval
of field sweep times
\begin{equation}\label{eq:validity1}
\frac{h_0}{(h_0 - \omega_0)^2} \ll \tau_h \ll \tau_h^{(+)} \ ,
\end{equation}
where $\tau_h^{(+)}$ is a solution of
$$
\sqrt{\frac{\tau_h}{h_0}}
 e^{ - \alpha\tau_h
 \frac{(h_0 - \omega_0)^2}{4h_0}} = \frac{1}{ h_{\perp} }\ .
$$

\begin{figure}[b]
\center
\includegraphics[width=0.48\textwidth]{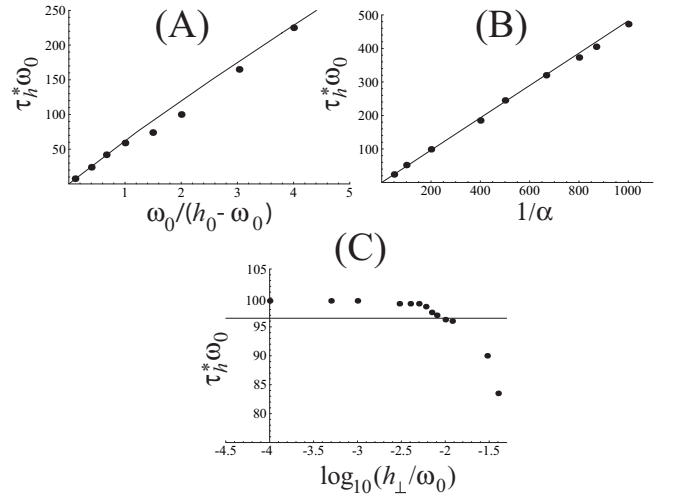}
\caption{Numeric (points) and approximate analytical (solid lines)
dependencies of the optimal field sweep time $\tau_h^*$ on the
system parameters: (a) fixed $\alpha$ and $h_{\perp}$, (b) fixed
$h_0/\omega_0$ and $h_{\perp}$, (c) fixed $h_0/\omega_0$ and
$\alpha$. When not varied, the parameter values are $h_0 = 2.2
\omega_0$, $\alpha = 0.01$, $h_{\perp} = 0.005$. }
 \label{fig:taustar}
\end{figure}

The optimal sweep time $\tau_h^*$ is determined from
$\partial\tau_m/\partial\tau_h = 0$. We get an expression
\begin{equation}\label{eq:optimal_tauh}
\tau_h^* = \frac{1}{2 \alpha (h_0 - \omega_0)}\frac{4 h_0 }{3 h_0 +
\omega_0} \ .
\end{equation}
This formula is our second main result. Note that $\tau_h^*$ is
independent of the bias field. The minimal switching time
$\tau_m(\tau_h^*)$ itself depends on $h_{\perp}$, which is quite
natural since the initial deviation from the easy axis is controlled
by $h_{\perp}$. We have also calculated the switching time drop
$$
\tau_m(0) - \tau_m(\tau_h^*) = \frac{\ln \left(
    \frac{\pi (h_0 - \omega_0)(h_0 + \omega_0)^2}{2 \alpha h_0^2 (3 h_0 + \omega_0)}
  \right) - 1}{2 \alpha (h_0 - \omega_0)} \ .
$$
between the instant and the optimal field sweeps. The drop is
independent of the bias field as long as approximation
(\ref{eq:tau_m_preview}) is valid.

Formula (\ref{eq:optimal_tauh}) is valid when $\tau_h^*$ falls into
the interval (\ref{eq:validity1}). Our calculations show that this
is guaranteed for
\begin{equation}\label{eq:alpha_requirement}
\frac{h_{\perp}^2}{h_0(h_0-\omega_0)} \ll \alpha \ll \frac{h_0 -
\omega_0}{2 h_0} \ .
\end{equation}
These inequalities place a more stringent constraint on the Gilbert
damping than the simple $\alpha \ll 1$.

Fig.~\ref{fig:alpha_family} compares numerically calculated
switching times with our analytic formula (\ref{eq:tau_m_preview}).
When inequalities (\ref{eq:alpha_requirement}) are well satisfied,
the quality of approximation is very good. As one approaches the
limits of the approximation's validity by, e.g., increasing
$\alpha$, the errors grow larger.

Fig.~\ref{fig:taustar} shows the dependence of the optimal field
sweep time $\tau_h^{*}$ on the system parameters. The correspondence
with formula (\ref{eq:optimal_tauh}) is generally good, although
some visible deviations exist. The accuracy of the determination of
$\tau_h^*$ is lowered by a flat shape of the $\tau_m(\tau_h)$
minimum. The shallow minimum, however, also lowers the practical
importance of precise determination of $\tau_h^*$.

In general, the analytic expression can approximate the
$\tau_m(\tau_h)$ dependence up to a 10\% accuracy in a surprisingly
wide range of parameters. Such accuracy is certainly sufficient for
the estimates related to the device design.

\begin{figure}[t]
 \center
\includegraphics[width = 0.27 \textwidth]{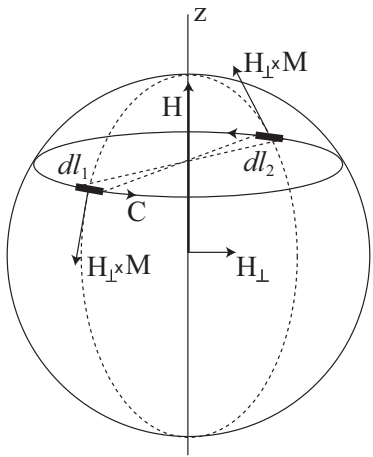}
\caption{Average ``ballistic'' contribution of the bias field.
Vector ${\bf M}(t)$ orbits around a parallel circle $C$ on a sphere
$|{\bf M}| = M_0$. The torque due to ${\bf H}_{\perp}$ pushes $\bf
M$ along the meridians of the sphere. In constant switching field
$\bf H$ the torque contributions from the diametrically opposed
elements $dl_1$ and $dl_2$ cancel each other. For variable ${\bf
H}(t)$ such a cancelation does not happen because ${\bf M}$ spends
unequal amounts of time on $dl_1$ and $dl_2$.}
 \label{fig:ballistic_contribution_averaging}
\end{figure}

We now discuss the physical reason for the minimum of the function
$\tau_m(\tau_h)$. The bias field has two roles in the switching
process. First, it provides the initial deviation from the easy
axis. Second, it alters the equations of motion for ${\bf M}(t)$.
The derivation of Eqs.~(\ref{eq:tau_m_preview}) and
(\ref{eq:optimal_tauh}), which will be detailed in our forthcoming
paper, shows that the second contribution is dominant. Recall now
that in the absence of anisotropy and other fields the torque ${\bf
H}_{\perp} \times {\bf M}$ due to the bias field would rotate vector
$\bf M$ from $+\hat z$ to $- \hat z$ along a meridian of the sphere
$|{\bf M}| = M_0$ (dashed line in
Fig.~\ref{fig:ballistic_contribution_averaging}), in a ballistic
(precessional) fashion. In our case a weak bias field is applied on
top of the strong uniaxial anisotropy and switching field, which
together induce a fast orbital motion of vector ${\bf M}(t)$ along
the parallel circles (line $C$ in
Fig.~\ref{fig:ballistic_contribution_averaging}). The bias field
still attempts to move $\bf M$ along the meridians, but now its
action has to be averaged over the orbital period. As illustrated in
Fig.~\ref{fig:ballistic_contribution_averaging}, in constant fields
${\bf H} = \pm H_0 \hat z$ averaging gives zero due to the
cancelation of the contributions from  the diametrically opposed
infinitesimal intervals $dl_1$ and $dl_2$ of equal lengths. This way
ballistic contribution of the bias field is quenched. However, the
contribution of ${\bf H}_{\perp}$ does not average to zero for a
variable switching field ${\bf H}(t)$. In this case the velocity of
$\bf M$ changes along the orbit, the times spent in the intervals
$dl_1$ and $dl_2$ are different, and the contributions of the two do
not cancel each other. We conclude that in the presence of a time
dependent external field ${\bf H}(t)$ ballistic contribution of the
perpendicular bias field is recovered. Moreover, this contribution
helps to move vector $\bf M$ from $+ \hat z$ to $- \hat z$ and is
thus responsible for the initial decrease of $\tau_m$. As the sweep
time grows larger, the change of the orbital velocity during the
precession period decreases and the ballistic contribution averages
out progressively better. The helping effect of ballistic switching
is lost and $\tau_m$ starts to increase as it normally would.

Ballistic contribution to switching can be also viewed as a
phenomenon complimentary to the magnetic resonance and rf-assisted
switching,\cite{thirion:2003,bertotti:2001,rivkin:2006} where ${\bf
H}$ is constant but ${\bf H}_{\perp}(t)$ is time-dependent. There
the average contribution of the bias field on an orbit does not
vanish due to the time dependence of ${\bf H}_{\perp}$. The
non-vanishing contribution, regardless of its origin, assists the
switching and makes it faster.

Our analytical results provide a convenient approximation for the
optimal field sweep time, an important parameter in the device
design. They can be used as a starting point for the investigations
of the switching time in granular media, where each grain can be
modeled by a single moment and bias fields are produced by the other
grains or by the spread of grain orientations.

Ya.~B. Bazaliy is grateful to B. V. Bazaliy for illuminating
discussions. This work was supported by the NSF grant DMR-0847159.


\begin{thebibliography}{2}

\bibitem{back:1999}
C. H. Back, R. Allenspach, W. Weber, S. S. P. Parkin, D. Weller, E.
L. Garwin, and H. C. Siegmann, Science {\bf 285}, 864 (1999).

\bibitem{xiao:2006}
Di Xiao, M. Tsoi, and Qian Niu,
%Minimal field requirement in precessional magnetization switching
J. Appl. Phys. {\bf 99}, 013903 (2006).

\bibitem{wang:2007}
X. R. Wang and Z. Z. Sun,
%Theoretical limit in the magnetization reversal of Stoner particles
Phys. Rev. Lett. {\bf 98}, 077201 (2007).

\bibitem{stankiewicz}
A. Stankiewicz, APS March Meeting, Portland, OR, March 15-19, 2010.
Bulletin of the APS {\bf 55}(2), abstract H33.11.

\bibitem{suess:2002}
D. Suess, T. Schrefl, W. Scholz, and J. Fidler,
J. Magn. Magn. Mater. {\bf 242-245}, 426 (2002).

\bibitem{he:1994}
L. He, W. D. Doyle, and H. Fujiwara, IEEE Trans. Magn. {\bf 30},
4086 (1994).

\bibitem{porter:1998}
D. G. Porter, IEEE Trans. Magn. {\bf 34}, 1663 (1998).

\bibitem{thirion:2003}
C. Thirion, W. Wernsdorfer, and D. Mailly,
Nat. Mater. {\bf 2}, 524 (2003).

\bibitem{bertotti:2001}
G. Bertotti, C. Serpico, and I. D. Mayergoyz,
Phys. Rev. Lett. {\bf 86}, 724 (2001).

\bibitem{rivkin:2006}
K. Rivkin and J. B. Ketterson,
Appl. Phys. Lett. {\bf 89}, 252507 (2006).
\end{thebibliography}
\end{document}